\documentstyle[preprint,aps,prl]{revtex} 
\begin{document} 

\title{ Crossover of Level Statistics between Strong 
and Weak Localization in Two Dimensions } 

\author{Isa Kh.~Zharekeshev$^1$, Markus Batsch$^2$ and Bernhard Kramer$^1$}

\address{
$^1$I. Institut f\"ur Theoretische Physik, Universit\"at Hamburg,
Jungiusstrasse~9, D-20355 Hamburg, Germany\\
$^2$Physikalisch-Technische Bundesanstalt, Bundesallee 100, D-38116
Braunschweig, Germany\smallskip\\
\rm{(February 2, 1996)}\bigskip\\
\parbox{14.2cm}{\rm
We investigate numerically the statistical properties of spectra 
of two-dimensional disordered systems 
by using the decimation method applied to the Anderson model.
Statistics of the spacings calculated for system size up to
 $1024 \times 1024$ lattice sites exhibits a crossover between Wigner
and Poisson distributions.
We perform a self-contained finite-size scaling 
analysis to find a single-valued
one-parameter function $\gamma (L/\xi )$ which governs the
crossover. 
The scaling parameter $\xi(W)$ is deduced and compared with the
localization length.      
$\gamma (L/\xi )$ does {\em not} 
show
critical behavior
and has two asymptotic regimes corresponding to 
weakly and strongly localized states.
}}

\maketitle
\narrowtext

\newpage 

The single-parameter scaling theory of
Anderson localization suggests that there exists no conducting state 
in an infinite two-dimensional (2D) disordered system at zero 
temperature~\cite{AALR}, since all one-electron wave functions 
are localized even for arbitrarily small fluctuations of a random 
potential~\cite{Lee85}.
The corresponding energy levels are distributed completely randomly.
One expects that there are no spectral correlations at {\em any} energy scale.
Therefore,
for any non-vanishing disorder 
the distribution of neighboring level spacings $P(s)$ 
is given by the Poisson law $P_{P}(s) = \exp(-s)$
in the thermodynamic limit. 
(Energy is measured in units of the mean level spacing $\Delta$.)

For finite systems $P(s) \approx P_{P}(s)$
only if the disorder $W$ is strong enough such that 
the localization length $\xi$ is shorter than the system size $L$.
Indeed, for fixed $W\ne 0$
the level spacing distribution 
was numerically shown to approach
$P_{P}(s)$
with increasing $L$~\cite{decimation}.
With decreasing $W$
the electron states change
gradually from ``strongly localized'' to ``weakly localized'', 
and the energy levels become more correlated.
When $\xi \gg L$, their statistics is well described by the Wigner-Dyson 
theory for eigenvalues of random matrices~\cite{RMT}.

As a consequence
 of the transition from strong to weak localization, 
$P(s)$ is expected to undergo a continuous crossover
between 
$P_{P}(s)$, and the Wigner surmise 
$P_{W}(s)=\pi s/2  \exp(-\pi s^2/4)$.
In the weakly localized and non-diffusive regimes the spectral
fluctuations were studied in terms of the two-level correlation
function by the use of 
sophisticated analytical techniques~\cite{Kravtsov,AltlandG95}.
Small corrections to the Wigner-Dyson statistics were obtained in~\cite{mirlin}.
Although the absence of the metal-insulator transition 
for statistics of the spectra in 2D-system was already mentioned 
by Shklovskii {\em et al}~\cite{shklovskii},
so far extensive investigations in the intermediate regime, where
the deviations from both of the limiting distributions are considerable, 
have not yet been 
carried out. 
An important question is which scaling law governs
the crossover of the level statistics.
For three-dimensional (3D) disordered systems such a scaling law 
shows critical behavior at a 
finite disorder which is the signature of a metal--insulator 
transition~\cite{shklovskii,hofstetter,isa}.
If no delocalization transition exists in 2D~\cite{remark}, $P(s)$
should {\it not} exhibit critical behavior at any finite disorder.
In this paper we provide 
strong numerical evidence that this is indeed the case.

The scaling behavior of $P(s)$ with $L$ and $W$ 
is determined 
by using the results for the energy spectrum that were obtained by 
a combination of the decimation method and a Lanczos procedure. 
Due to the achieved large 
system sizes, 
the scaling analysis could be performed 
very accurately. 
{\em Independently} of other methods,
we calculate a one--parameter
scaling variable $\gamma (L/\xi )$ related to $P(s)$ 
which governs a crossover between the two above limits. 
It turns out to be monotonic without
any critical
point $W_{c}$, for which $\gamma(L,W_{c})$ = const,
that would indicate a delocalization transition.
A similar variable $\gamma (L/\xi_{TMM})$ 
was claimed in~\cite{shklovskii} to 
behave as a single-valued function, where $\xi_{TMM}$ is the 
localization length obtained from the transfer-matrix method 
(TMM)~\cite{mackinnon}.  
We find that 
in the strongly localized regime the deviation from
the Poissonian
is described by a linear law
$\gamma \propto L/\xi$.
In the weakly localized regime $\gamma$ 
tends towards
the Wigner limit.
From spectral statistics we determine 
the scaling parameter $\xi$,
which can be interpreted as the localization length. 
Its dependence on $W$ is consistent with the results 
of previous calculations using the TMM~\cite{mackinnon,schreiber}.

We consider the Anderson model (AM) with diagonal disorder
$ H = \sum_{i} \varepsilon_i \mid i \ \rangle \langle \ i \mid + t \sum_{<i,j>}
 \mid i \ \rangle \langle \ j \mid $,
where $i$ labels the $L^2$ sites of a 
square lattice of the 
linear size $L$. The on-site energies $\varepsilon_i$
are independently distributed  
at random within an interval of width $W$.
The second sum is taken over all pairs of nearest-neighbor sites. 
It corresponds to the kinetic energy.  
For the calculations we used a decimation
algorithm described in~\cite{decimation,lee}, which enables us to investigate 
the level statistics 
near the band center $E_0$  for systems 
containing up to 
$10^5$--$10^6$ lattice sites.
The algorithm consists of a sequence of iterative steps~\cite{lee}, 
in which 
the spectrum of a system
is computed by using the amplitudes of wave functions and
corresponding eigenvalues 
of four smaller systems of 
half of the linear size. 
Each step 
consists of 
(i) a unitary transformation of the secular matrix to the 
basis set of the smaller lattices, 
(ii) a truncation of the corresponding  Hilbert space to a
fixed number $m$ of states 
near $E_0=0$,
and 
(iii) a diagonalization of the truncated Hamiltonian. 
The iteration starts with the original 2D lattice of the AM. 
The latter is divided into many
independent small square cells of linear size $L_{o}$, which is defined by 
the disorder.
Subsequent larger squares are constructed by
coupling four smaller squares via hopping elements $t$
only between corresponding boundary sites.
In the last step of the iteration, periodic boundary
conditions are imposed. 
Using ensembles of up to $k \approx 1000$ different realizations of the 
random potential, we calculate $P(s,L,W)$ at $E_{0}$. 
                              
As expected, we find the Poisson distribution for large $L$ and large 
$W$ and the Wigner distribution for small $W$ and small $L$.
The 
crossover between the two 
limiting cases can be detected for fixed 
disorder with increasing $L$, or for fixed $L$ with increasing
$W$.
Figure~\ref{fig1} shows 
that for $W=4$ 
the level spacing distribution 
is only  slightly different from the Wigner surmise. 
On the other hand, $P(s)$ for $W=7$
already resembles strongly the Poisson distribution.
This can be explained by the 
considerably different values of the localization length for these two 
disorders. 
In the former case, $\xi \gg L$ 
while in the latter case 
$\xi \ll L$
(see the inset of Fig. 3 below).
In order to test our method we also calculated $P(s)$
by using 
a Lanczos procedure~\cite{lanczos}, 
especially optimized for large sparse
matrices~\cite{isa}. We were able to compute the exact 
spectrum for systems
up to $L^2=256^2$. 
The results of the decimation method for $P(s)$
are in good agreement 
with those 
obtained by the exact diagonalization 
(inset of Fig.~\ref{pofs}).
For larger systems the Lanczos procedure failed even for a very narrow
energy interval due to limitations of computing time, while we could enlarge
the system up to $L^2\sim 10^6$ lattice sites by 
using the decimation algorithm.
The striking advantage of the latter is that it allows to go
beyond the limits 
of conventional
diagonalization procedures,
at the expense of being 
restricted to a 
small (and constant) number of energy levels around the 
energy of interest.

The decimation scheme is applicable as long as $V \ll \Delta$, where
$V$ is the typical matrix element between states of
the coupled blocks (see~\cite{decimation,lee}). 
Therefore one should start the calculations with such $L_{o}$ for a given 
$W$, so that this condition is fulfilled. 
The choice of $m$ defines the relative error of the retained eigenvalues 
$\delta E/\Delta \sim V^2/(\Delta^2 m) \ll 1$~\cite{lee},
resulting from the neglection of the discarded levels.
The errors of the consecutive iterations steps are mostly dominated 
by those of the first step, which diminish with decreasing coupling, 
i.e. with increasing $W$.
In order to control the precision of $P(s)$  we
choose $L_{o}$ and $m$ depending on $W$ so that 
after several iterations $V/\Delta \lesssim 0.01$. 
The initial size grows from $L_{o}=16$ for $W \ge 8$ to $L_{o}=256$ for 
$W\le 5$
and $m$ ranges typically from 25 to 100.
For the sizes smaller
than $L_{o}$ the exact diagonalization is applied.

In order to investigate the scaling behavior of the level statistics 
quantitatively,
we consider the integral
$I(s) \equiv \int_{0}^{s} P(s')ds'$.
It describes the probability to have a spacing smaller than $s$.
Numerically, $I(s,L,W)$ can be obtained with a higher accuracy than 
$P(s,L,W)$.
Figure~\ref{int} shows
how $I(s)$ changes from 
$I_{W}(s)=1-\exp(-\pi s^2 /4)$ to $I_{P}(s) = 1 - \exp(-s)$
for fixed disorder in approaching the thermodynamic limit.
For example, the small-$s$ behavior of $I(s,L,W)$ changes from quadratic
for small to  
linear 
for larger system sizes.  Also, the probability to find
large spacings is enhanced, when increasing $L$.
 
 The next step 
is to determine 
the deviation of the level statistics
from the Poissonian limit due to the finiteness of the system.
We introduce the function
\begin{equation}
\gamma (L, W) \equiv
\frac{1}{N} \int_{0}^{s_{0}} \left[ I(s,L,W) - I_{P}(s) \right] ds,
\label{gamma}
\end{equation}
which is similar to the quantities considered 
in Ref.~\cite{shklovskii,hofstetter,isa}.
The normalization constant $N$ is chosen such that        
$\gamma=1$, if $I(s) =I_{W}(s)$. For the scaling analysis 
the choice of the upper limit $s_{0}$ of the integral 
does not play an important role~\cite{isa}.
By optimizing the numerical accuracy we fixed it to $s_{0} = 1.25$$\Delta$.
For an 
infinite 2D disordered system 
$\gamma = 0$ for all $W\ne 0$.
For finite systems, $\gamma$ is 
expected to change continuously from unity to zero 
with increasing $L$~\cite{rem2}.

We computed $\gamma$ for 
$W$ from 4 to 12 and $L$ from 8 to 1024
(inset of Fig.~\ref{int}).
One observes that 
$\gamma$ decreases with increasing 
$L$ 
for all $W$. 
For weak disorder it is close to unity 
and decreases slowly, as the system doubles in size,
while for larger 
disorder the
decrease is 
more rapid. 
Correspondingly, for a fixed $L$, $\gamma$ diminishes as the disorder
becomes stronger.
The data for $W \gtrsim 7$ 
saturate towards $\gamma = 0$. 
Altogether, $\gamma(L,W)$ undergoes
a smooth monotonic crossover from unity to zero as $L$ grows,  
with no critical behavior, 
at least for
the disorders considered here.
This is markedly different from the 3D case, where a critical disorder 
$W_{c}\ne 0$ exists,
below which
the electron states are delocalized, and 
$\gamma $ should increase with $L$~\cite{shklovskii,isa}. 
We have estimated the error of $\gamma$, $\delta \gamma /\gamma \approx
(\delta \gamma /\gamma)_{stat} + (\delta \gamma /\gamma)_{syst}$.
The former is of order of  $(k m I(s_{o}))^{-1/2} < 2 \% $.
The latter can be roughly 
estimated as $I(s_{o})/(N \gamma) \, \delta E/\Delta$ and is 
about $0.5 \%$ near $L_{o}/\xi \gtrsim 1$.
In the strong localization it decreases 
as $0.02\,L_{o}/\xi\,\exp(-L_{o}/\xi)$. 
Thus,  the statistical error due to the finite number of samples
is always larger than the systematic one resulting from the approximation.

Assuming that $\gamma$  
obeys a one--parameter scaling law $\gamma(L,W) = f(L/\xi)$, 
one can find the scaling parameter $\xi(W)$ such that 
all numerical data fall onto a common curve.
We fixed the data for $W=7$ 
and then shifted the data corresponding to other $W$
to the left and to the right in a logarithmic scale, until 
the set of data of a given $W$
overlapped for at least one value of $L$ with 
those corresponding to another $W$.
One can determine $\xi$ by this procedure 
up to a common numerical factor.
The latter corresponds to a shift of all of the data in the inset of 
Fig.~\ref{int} 
by the same amount along the $x$-axis.
We found this factor by adjusting $\xi$ 
at large $W$
to the localization length, which was earlier found
by using the TMM~\cite{schreiber}. 
For this fit
we choose $\xi_{TMM}(W=10)$ which
was smaller than the width of the quasi-1D strips used in 
the TMM calculations.
As shown in Fig.~\ref{scal},
%
%
%
the data of $\gamma(L/\xi)$ for
all pairs of $ \{ L, W \} $ collapse onto an one-branch curve within the 
numerical accuracy.
Thus, our calculations confirm by a completely independent procedure
that the crossover from the 
weak to the strong localization in 2D, 
which corresponds in the language of the 
level statistics to the crossover from 
the Wigner surmise to the 
Poissonian distribution,  
is governed by one-parameter scaling. 
The left, very flat part of the curve in Fig.~\ref{scal}, 
obtained for small disorders,
corresponds to the weak localization. Here, the data 
for $\ln \gamma $
approach zero because the scaling parameter $\xi(W)$
becomes large compared to the size of the system.
The small, but increasing deviation from 
$\gamma = 1$ for 
$L/\xi \ll 1$  
is in agreement 
with analytical results obtained earlier for the corrections to 
the Wigner-Dyson 
statistics in the case $s_{0}\sim \Delta$~\cite{mirlin}. 
Details of a quantitative comparison which involves also a calculation
of the conductance of the system are given elsewhere~\cite{batsch}.

When $L/\xi \gg 1$
the data are well described by
$\gamma = \gamma_{o} (\xi/L)$, where $\gamma_{o} \approx 1.9$.
An analogous linear behavior was found previously for the Lyapunov 
exponents~\cite{mackinnon}.
The fact that for strong localization the conductance
$g \propto \exp(-2 L/\xi)$ suggests that
$\gamma \propto -(\ln g)^{-1}$ for $L/\xi \gg 1$.
The intermediate part of the graph, 
where $L \sim \xi$, shows the smooth crossover between
the two limits.
Since $\gamma (L/\xi)$ consists of only a decaying 
branch, we conclude that there is no metal-insulator transition
at least for $W\geq 4$.
The inset of Fig.~\ref{scal} shows $\xi(W)$.
For $W<6$ one observes a faster increase of $\xi$ with decreasing $W$ 
than the localization length computed via TMM in~\cite{mackinnon}. 
On the other hand, we find good agreement with the values of $\xi_{TMM}$, 
reported recently in Ref.~\cite{schreiber}. 
It is reasonable to idendify our scaling parameter $\xi$ with the localization 
length of the electron states.
The disorder dependence of $\xi$
is consistent with the theoretical
prediction of an essential singularity  of the localization length 
at $W=0$~\cite{vollhardt} in the form 
$\xi (W) = a \exp [ (W_{o}/W)^{2}]$, where the length scale $a$ is of
order of the lattice constant.
By fitting the calculated data of $\xi $ 
to this formula we found  $a=1.1 \pm 0.25$ and $W_{o}=11 \pm 3$.

In conclusion, we performed a finite size scaling analysis
of the fluctuations in one-electron spectra of 2D disordered systems 
by using a combination of exact diagonalization and decimation algorithm. 
It enables us to treat systems up to $1024^{2}$ lattice sites.  
The nearest-neighbor level spacing statistics 
described by the variable $\gamma$~(Eq.~\ref{gamma}) 
indicates a smooth crossover from strong to weak localization 
when the disorder 
and the system size decrease.
The crossover is found to obey a one--parameter
scaling law which does not show critical behavior. This confirms the earlier 
conjecture~\cite{AALR}
that in 2D there are no true metallic states
in the thermodynamic limit. 
Our present results for the scaling parameter $\xi$ extracted from
the level statistics are quantitatively
consistent with those obtained
previously for the localization length 
by using a completely different numerical approach
\cite{schreiber}, and
conclusions based on perturbative calculations~\cite{vollhardt}. 

We thank L. Schweitzer and B. Altshuler for helpful discussions.
Financial support from SCIENCE program, grant SCC$^\ast$-CT90-0020, 
HCM program, grant CHRX-CT93-0136 
and Deutsche Forschungsgemeinschaft is 
gratefully acknowledged. We also thank the HLRZ
at KFA J\"ulich for providing CPU-time.

\newcommand{\pap}[5]{#1,\ #2{\bf #3}, #4 (19#5)}

\begin{figure}[p]
\caption[pofs]{\label{pofs}The probability density of neighboring
spacings $P(s)$ for system size $L^2=256^2$ 
at different disorders. 
Solid and dotted lines: Wigner surmise and Poisson law, respectively. 
Inset: $P(s)$ 
at $W=6$ obtained by the decimation and 
the Lanczos method.} 
\label{fig1} 
\end{figure} 
\begin{figure}[p]
{\caption[int]{\label{int} Total probability distribution of spacings
$I(s)$ at the disorder $W$=6 for various sizes. 
Solid and dotted lines: $I_{W}(s)$ and $I_{P}(s)$,
respectively. 
Inset: $\gamma(L)$ 
for different disorders $W=4.0 \ (\ominus)$, $4.5 \ (\star)$, $5.0 \ (\circ)$
, $5.5 \ (+)$, $6.0 \ (\bullet)$, $7.0 \ (\triangle) $, $8.0 \ ( \times )$, 
$9.0 \ (\Box )$, $10.0 \ (\bigtriangledown)$, $12.0 \ (\diamond) $.
}} \end{figure} 
\begin{figure}[p]
{\caption[scal]{\label{scal} The parameter $\gamma$ as a function of 
$L / \xi(W)$. Straight line: limit of strong 
localization. 
Symbols for $W$ are the same as in the inset of Fig.~\ref{int}.
Inset: disorder dependence of the scaling parameter $\xi$ ($\bullet$) 
along with the results obtained for the localization length
by TMM~\cite{mackinnon} ($\times$) and~\cite{schreiber} ($\bigtriangleup$). 
}} \end{figure} 


\begin{thebibliography}{99}
\bibitem{remark} Though one-parameter scaling theory suggests this, there is 
no mathematically rigorous proof for the absence of extended states in 2D, 
yet.
\bibitem{rem2} In fact the limit $\gamma=0$ is reachable unless
the electron mean free time exceeds $\Delta^{-1}$ 
(see Ref.~\cite{AltlandG95}).
\end{thebibliography}
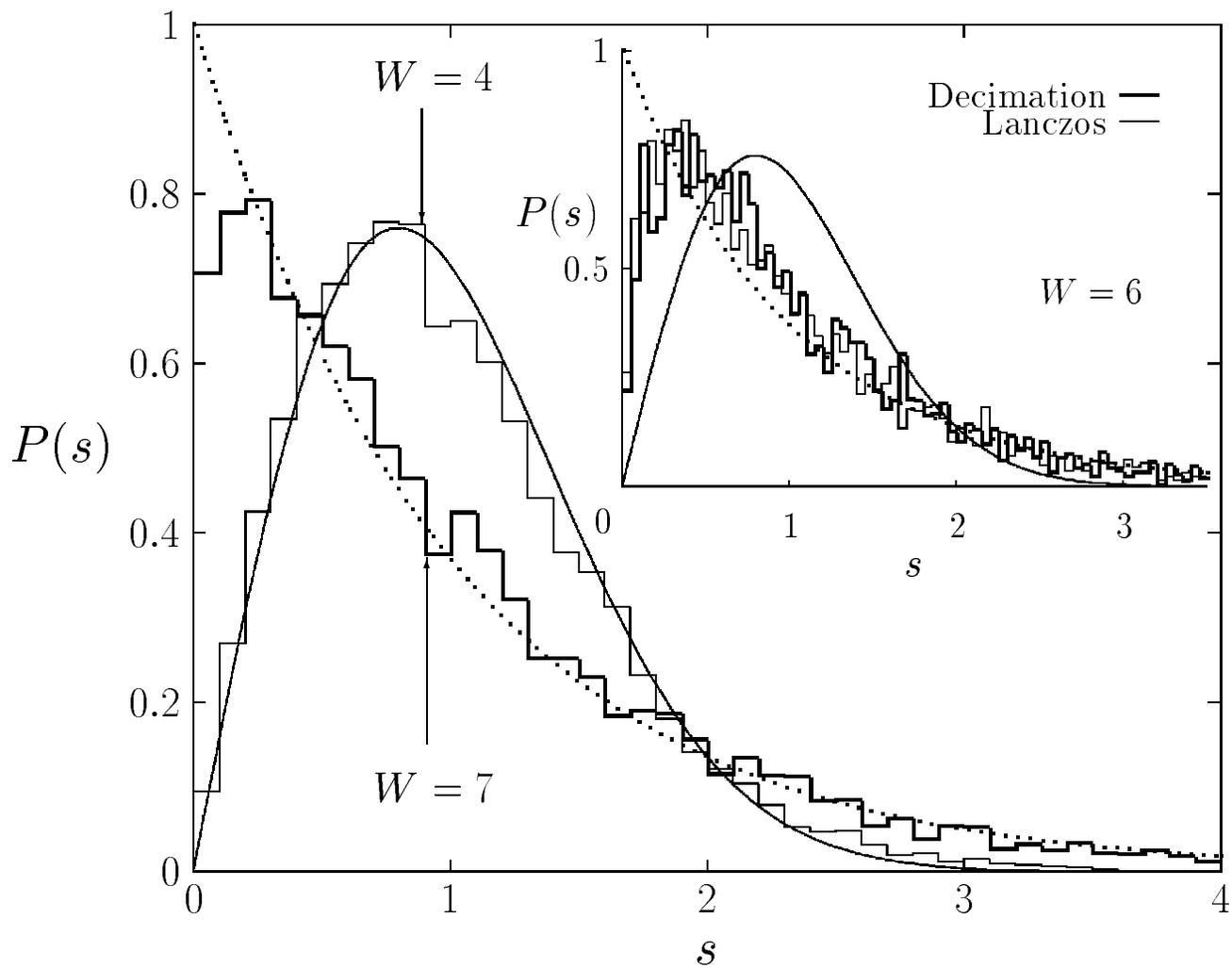
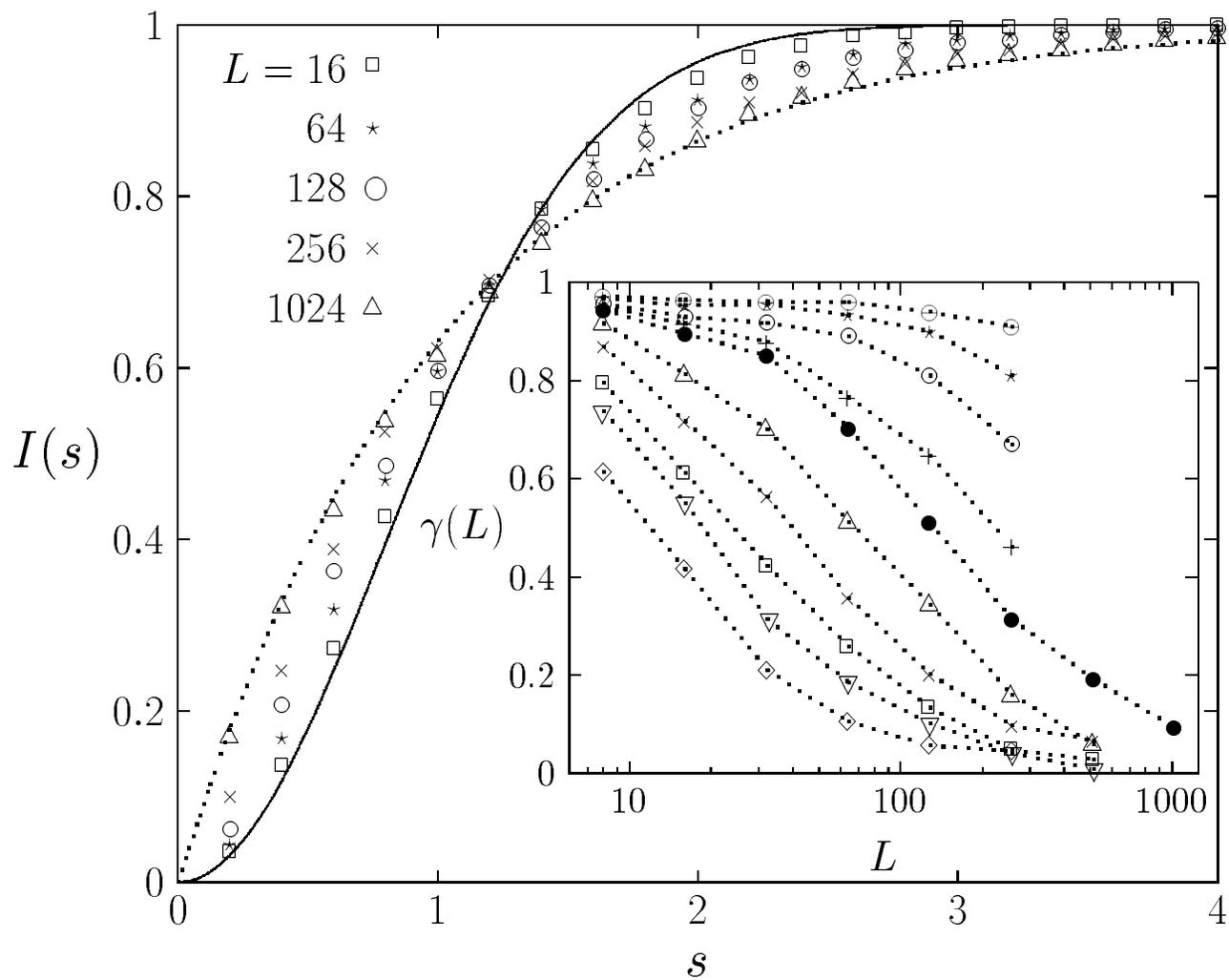
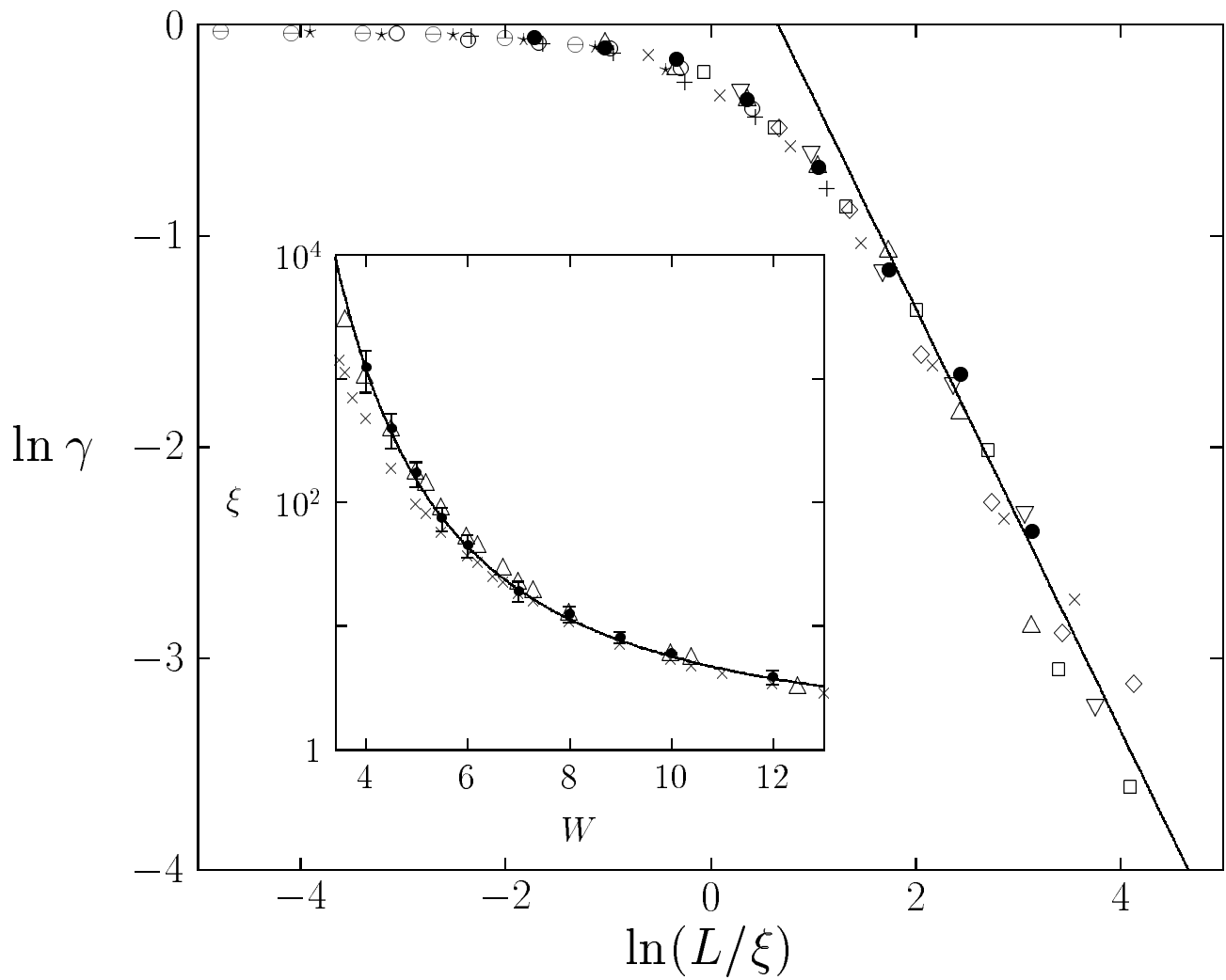
\end{document}